\PassOptionsToPackage{unicode=true}{hyperref} 
\PassOptionsToPackage{hyphens}{url}

\documentclass[a4paper, 12pt]{article}

\usepackage[sort&compress]{natbib}
\bibpunct{(}{)}{;}{a}{}{,} 

\usepackage{amsthm, amsmath, amssymb, mathrsfs, multirow, url, subfigure}
\usepackage{graphicx} 
\usepackage{ifthen} 
\usepackage{amsfonts}
\usepackage[usenames]{color}
\usepackage{float}

\usepackage[margin=1in]{geometry}
\usepackage{fancyhdr}
\theoremstyle{plain}

\theoremstyle{definition}

\theoremstyle{remark}

\IfFileExists{upquote.sty}{\usepackage{upquote}}{}
\IfFileExists{microtype.sty}{%
\usepackage[]{microtype}
\UseMicrotypeSet[protrusion]{basicmath} 
}{}
\IfFileExists{parskip.sty}{%
\usepackage{parskip}
}{
\setlength{\parindent}{0pt}
\setlength{\parskip}{6pt plus 2pt minus 1pt}
}
\usepackage{hyperref}
\hypersetup{
            pdftitle={Event based simulation of an EPR-B experiment by local hidden variables: epr-simple and epr-clocked},
            pdfauthor={Richard D. Gill},
            pdfborder={0 0 0},
            breaklinks=true}
\usepackage[margin=1in]{geometry}
\usepackage{color}
\usepackage{fancyvrb}

\DefineVerbatimEnvironment{Highlighting}{Verbatim}{commandchars=\\\{\}}
\usepackage{framed}
\definecolor{shadecolor}{RGB}{248,248,248}
\newenvironment{Shaded}{\begin{snugshade}}{\end{snugshade}}

\newcommand{\CommentTok}[1]{\textcolor[rgb]{0.56,0.35,0.01}{\textit{#1}}}

\newcommand{\ControlFlowTok}[1]{\textcolor[rgb]{0.13,0.29,0.53}{\textbf{#1}}}
\newcommand{\DataTypeTok}[1]{\textcolor[rgb]{0.13,0.29,0.53}{#1}}
\newcommand{\DecValTok}[1]{\textcolor[rgb]{0.00,0.00,0.81}{#1}}

\newcommand{\FloatTok}[1]{\textcolor[rgb]{0.00,0.00,0.81}{#1}}

\newcommand{\KeywordTok}[1]{\textcolor[rgb]{0.13,0.29,0.53}{\textbf{#1}}}
\newcommand{\NormalTok}[1]{#1}
\newcommand{\OperatorTok}[1]{\textcolor[rgb]{0.81,0.36,0.00}{\textbf{#1}}}
\newcommand{\OtherTok}[1]{\textcolor[rgb]{0.56,0.35,0.01}{#1}}

\newcommand{\StringTok}[1]{\textcolor[rgb]{0.31,0.60,0.02}{#1}}

\usepackage{graphicx,grffile}
\makeatletter
\def\maxwidth{\ifdim\Gin@nat@width>\linewidth\linewidth\else\Gin@nat@width\fi}
\def\maxheight{\ifdim\Gin@nat@height>\textheight\textheight\else\Gin@nat@height\fi}
\makeatother
\setkeys{Gin}{width=\maxwidth,height=\maxheight,keepaspectratio}
\ifx\paragraph\undefined\else
\let\oldparagraph\paragraph
\renewcommand{\paragraph}[1]{\oldparagraph{#1}\mbox{}}
\fi
\ifx\subparagraph\undefined\else
\let\oldsubparagraph\subparagraph
\renewcommand{\subparagraph}[1]{\oldsubparagraph{#1}\mbox{}}
\fi

\makeatletter
\def\fps@figure{htbp}
\makeatother

\title{Event-based and LHV simulation of an EPR-B experiment:
\texttt{epr-simple} and \texttt{epr-clocked}}
\author{Richard D. Gill\footnote{Mathematical Institute, Leiden University, Netherlands}
\\ \small  \url{https://www.math.leidenuniv.nl/~gill}
\\ \small 
This is version 11 of this preprint. Earlier versions incorrectly stated \\ \small 
that documentation of Michel Fodje's simulation was inexistent. This \\ \small 
version also contains some further explanations of my terminology.}
\date{April 10, 2021}

\begin{document}
\maketitle
\thispagestyle{fancy}
\begin{abstract}    
In this note, I analyse the data generated by M. Fodje's
(2013, 2014) simulation programs ``\texttt{epr-simple}'' and
\lq\lq\texttt{epr-clocked}\rq\rq. They are written in Python and were
published on \textit{Github}. 
Inspection of the program descriptions shows that they make use of the
detection-loophole and the coincidence-loophole respectively.

I evaluate them with appropriate modified Bell-CHSH type inequalities:
the Larsson detection-loophole adjusted CHSH, and the Larsson-Gill
coincidence-loophole adjusted CHSH. The experimental efficiencies turn
out to be approximately \(\eta = 81\%\) (close to optimal) and
\(\gamma = 55\%\) (far from optimal). The observed values of CHSH are,
as they must be, within the appropriate adjusted bounds. Fodjes'
detection-loophole model turns out to be very, very close to Pearle's
famous 1970 model, so the efficiency is very close to optimal. The model
also has the same defect as Pearle's: the joint detection rates exhibit
signalling.

Fodje's coincidence-loophole model is actually an elegant modification of
his detection-loophole model. However, this particular construction prevents 
attainment of the optimal efficiency.
\end{abstract}

\hypertarget{introduction}{%
\section{Introduction}\label{introduction}}

Michel Fodje, in 2013--2014, wrote two event-based simulation programs
of EPR-B experiments called \lq\lq\texttt{epr-simple}\rq\rq\, and
\lq\lq\texttt{epr-clocked}\rq\rq. The programs are written in the Python
programming language and are freely available at
\url{https://github.com/minkwe/epr-simple} and
\url{https://github.com/minkwe/epr-clocked}. Descriptions are given at
\url{https://github.com/minkwe/epr-simple/blob/master/README.md} and
\url{https://github.com/minkwe/epr-clocked/blob/master/README.md}.

One can recognise from the program descriptions
that the two programs respectively use the detection loophole (Pearle, 1970) 
and the coincidence loophole (Pascazio, 1986; Larsson and Gill, 2004) to
reproduce (to a good approximation) the singlet state correlation
function and thereby to violate the CHSH inequality. See the appendix
for further information on the models. I initially imagined that the
author was perhaps inspired by event-based simulations, based on these
loopholes, which at that time had recently been published by Hans de
Raedt and his collaborators, and were being discussed in various
internet fora. However,  he had come up with the ideas behind these models, as well
as the particular implementation of the general ideas, completely by
himself. He never studied the literature on these loopholes, or even
considered them as \lq\lq\emph{loopholes}\rq\rq. He just tried,
independently, and successfully, to create local realistic event-based
simulation models which reproduced most features of past experiments. He
was certainly provoked by some remarks of mine. On an internet forum I
wrote, and he quoted these words,

\begin{quote}
It is impossible to write a local realist computer simulation of a \emph{clocked} experiment with no 
\lq\lq\emph{non-detections}\rq\rq, and which reliably reproduces the singlet correlations. (By reliably, I mean in the
situation that the settings are not in your control but are delivered to you from outside; the number 
of runs is large; and that this computer program does this not just once in a blue moon, by luck, but
most times it is run on different people's computers.)
\end{quote}

Unfortunately, I did not explain in the same internet posting exactly what I meant by the word
\lq\lq\emph{clocked}\rq\rq (the emphasis was mine). It would have been better if I had 
used the word ``\emph{pulsed}''. I was thinking of those experiments in which the time axis
is split into a regular sequence of fixed time intervals often called time slots, and in
each time slot and in each wing of the experiment only one setting
is chosen, at random, at the start of the time slot; then arbitrary algorithms in each
wing of the experiment generate one binary outcome from the data
which is collected there, and call it \emph{the outcome} \dots~even if
there is no detection event there at all. With regularly spaced time slots and pulsed
lasars this arrangement is nowadays quite standard. 

If there is no detection event at all in one time slot they should use a fixed
outcome, or even toss a coin to make one. Physicists (experimental or
theoretical) tend to howl with disapproval at this suggestion. But I was
serious. In medical statistics, this is called using the
\lq\lq\emph{intention to treat}\rq\rq\, principle, and is a standard way
to deal with non-compliance in double-blind randomised clinical trials.

Fodje wrote that his simulations do not use post-selection or use the detection loophole since
all emitted particles are detected.  He did not refer to Larsson and Gill (2004), and apparently
did not know the literature well, hence did not have any
idea what I meant by the word \lq\lq\emph{clocked}\rq\rq, though I
certainly explained what I meant many times on the internet fora which we
both frequented.  I meant that the experiment should be set up \emph{and} 
analysed as a long regular sequence 
or \emph{run} of $N$ \emph{trials}, each trial corresponding to a pair of time slots, one in each wing of the
experiment. Per trial, in a CHSH-Bell inequalities type experiment,
two random binary settings go in and two binary outcomes come out.
In other experiments, exploring the whole correlation function, the settings are angles, 
e.g., in whole numbers of degrees, chosen again and again, uniformly at random, in each wing of the experiment.
The outcomes are again binary.

Fodje's ``readme'' sections on Github include extensive
explanation of how the models work, but do not refer to
the now extensive literature on the two loopholes which Fodje effectively exploits.

I study the experimental efficiency of the two models in the CHSH
setting. For the detection loophole, the efficiency \(\eta\) is defined
to be the minimum over all setting pairs and over all permutations of
the two parties Alice and Bob, of the probability that Party 1 detects a
particle given that Party 2 has detected a particle. For the coincidence
loophole it is defined in a similar way: the efficiency \(\gamma\) is
defined to be the minimum over all setting pairs and over all
permutations of the two parties Alice and Bob, of the probability that
Party 1 has a detection which is paired to Party 2's detection, given
that Party 2 has detected a particle.

If either loophole is present in the experiment, then the CHSH
inequality is not applicable, or to be more precise, the statement that
local hidden variables cannot violate CHSH is not true. I refer the
reader to the survey paper Larsson (2014); arXiv eprint
\url{http://arxiv.org/abs/1407.0363}. One needs to make further
(untestable) assumptions such as the
\lq\lq\emph{fair sampling}\rq\rq\, hypothesis in order to deduce
impossibility of local hidden variables from violation of CHSH. However,
it is not difficult to modify CHSH to take account of the possibly
differential \lq\lq\emph{post-selection}\rq\rq of particle pairs which
is allowed by these two loopholes. The result is two bounds, replacing
the usual bound \lq\lq2\rq\rq: \(4/\eta - 2\) for the detection
loophole, and \(6/\gamma - 4\) for the coincidence loophole; see Larsson
(2014) formula (38) and formulas (50), (51), (52). Note that when
\(\eta = 1\) the detection loophole bound equals the usual CHSH bound 2,
but as \(\eta\) decreases from 1, the bound increases above 2, at some
point passing the best quantum mechanical prediction \(2\sqrt 2\) (the
Tsirelson bound) and later even achieving the absolute bound 4. The
bound is sharp: one can come up with local hidden variable models which
exactly achieve the bound at the given efficiency. In particular, with
\(\eta = 2/3\) the detection loophole bound is 4, saying that it is
possible for three of the four CHSH correlations to achieve their
natural upper limit $+1$ and one of them its lower natural limit $-1$.

The coincidence loophole bound is also attainable, and for the same
value of the efficiency, worse. In particular, already with
\(\gamma = 3/4\) one can attain three perfect correlations and one
perfect anti-correlation.

\hypertarget{epr-simple}{%
\section{epr-simple}\label{epr-simple}}

The programme epr-simple uses the detection loophole (Pearle, 1970) so
as to simulate violation of the CHSH inequality in a local-realistic
way. The simulated experiment can be characterised as a pulsed
experiment. At each of a long sequence of discrete time moments, two new
particles are created at the source, and dispatched to two detectors. At
the detectors, time and time again, a new pair of random settings is
generated. The two particles are measured according to the settings and
the outcome is either $+1$, $-1$, or $0$; the latter corresponding to
\lq\lq\emph{no detection}\rq\rq.

If either particle is not detected, the pair is rejected.

epr-simple only outputs some summary statistics for the accepted pairs.
I added a few lines of code to the program so that it also outputs the
\lq\lq\emph{missing data}\rq\rq. Not being an expert in Python
programming, my additional code is pretty simple.

First of all, I reduced the total number of iterations to 10 million.
The original code has 50 million, and this lead to memory problems on
the (virtual) Linux Mint system which I used for Python work. Secondly, I
added a code line \lq\lq\texttt{numpy.random.seed(1234)}\rq\rq in the
block of code called \lq\lq\texttt{class Simulation(object)}\rq\rq, so
that identical results are obtained every time I run the code. This
means that others should be able to reproduce the numerical results
which I analyse here, exactly.

Finally, in the part of the code which outputs the simulation results
for a test of the CHSH inequality, I added some lines to preserve the
measurement outcomes in the case either measurement results in
\lq\lq\emph{zero}\rq\rq\, and then to cross-tabulate the results.

By the way, for his test of CHSH, Michel Fodje (thinking of the
polarization measurements in quantum optics) took the angles 0 and 45
degrees for Alice's settings, and 22.5 and 67.5 for Bob. I have changed
these to 0 and 90 for Alice, and 45 and 135 for Bob, as is appropriate
for a spin-half experiment.

\begin{Shaded}
\begin{Highlighting}[]
        \ControlFlowTok{for}\NormalTok{ k,(i,j) }\ControlFlowTok{in} \KeywordTok{enumerate}\NormalTok{([(a,b),(a,bp), (ap,b), (ap, bp)])}\OperatorTok{:}
\StringTok{            }\NormalTok{sel0 =}\StringTok{ }\NormalTok{(adeg}\OperatorTok{==}\NormalTok{i) }\OperatorTok{&}\StringTok{ }\NormalTok{(bdeg}\OperatorTok{==}\NormalTok{j)              }\CommentTok{# New variable}
\NormalTok{            sel =}\StringTok{ }\NormalTok{(adeg}\OperatorTok{==}\NormalTok{i) }\OperatorTok{&}\StringTok{ }\NormalTok{(bdeg}\OperatorTok{==}\NormalTok{j) }\OperatorTok{&}\StringTok{ }
\StringTok{                     }\NormalTok{(alice[}\OperatorTok{:}\NormalTok{,}\DecValTok{1}\NormalTok{] }\OperatorTok{!=}\StringTok{ }\FloatTok{0.0}\NormalTok{) }\OperatorTok{&}\StringTok{ }\NormalTok{(bob[}\OperatorTok{:}\NormalTok{,}\DecValTok{1}\NormalTok{] }\OperatorTok{!=}\StringTok{ }\FloatTok{0.0}\NormalTok{)}
\NormalTok{            Ai =}\StringTok{ }\NormalTok{alice[sel, }\DecValTok{1}\NormalTok{] }
\NormalTok{            Ai0 =}\StringTok{ }\NormalTok{alice[sel0, }\DecValTok{1}\NormalTok{]                      }\CommentTok{# New variable}
\NormalTok{            Bj =}\StringTok{ }\NormalTok{bob[sel, }\DecValTok{1}\NormalTok{]}
\NormalTok{            Bj0 =}\StringTok{ }\NormalTok{bob[sel0, }\DecValTok{1}\NormalTok{]                        }\CommentTok{# New variable}
\NormalTok{            print }\StringTok{"%s: E(%5.1f,%5.1f), AB=%+0.2f, QM=%+0.2f"}\NormalTok{ % }
\NormalTok{                (DESIG[k], i, j, (Ai}\OperatorTok{*}\NormalTok{Bj)}\KeywordTok{.mean}\NormalTok{(),}
                 \OperatorTok{-}\KeywordTok{numpy.cos}\NormalTok{(}\KeywordTok{numpy.radians}\NormalTok{(j}\OperatorTok{-}\NormalTok{i)))}
\NormalTok{            npp =}\StringTok{ }\NormalTok{((Ai0 }\OperatorTok{==}\StringTok{ }\DecValTok{1}\NormalTok{) }\OperatorTok{&}\StringTok{ }\NormalTok{(Bj0 }\OperatorTok{==}\StringTok{ }\DecValTok{1}\NormalTok{))}\KeywordTok{.sum}\NormalTok{()     }\CommentTok{# New variable}
\NormalTok{            np0 =}\StringTok{ }\NormalTok{((Ai0 }\OperatorTok{==}\StringTok{ }\DecValTok{1}\NormalTok{) }\OperatorTok{&}\StringTok{ }\NormalTok{(Bj0 }\OperatorTok{==}\StringTok{ }\DecValTok{0}\NormalTok{))}\KeywordTok{.sum}\NormalTok{()     }\CommentTok{# New variable}
\NormalTok{            npm =}\StringTok{ }\NormalTok{((Ai0 }\OperatorTok{==}\StringTok{ }\DecValTok{1}\NormalTok{) }\OperatorTok{&}\StringTok{ }\NormalTok{(Bj0 }\OperatorTok{==}\StringTok{ }\DecValTok{-1}\NormalTok{))}\KeywordTok{.sum}\NormalTok{()    }\CommentTok{# New variable}
\NormalTok{            n0p =}\StringTok{ }\NormalTok{((Ai0 }\OperatorTok{==}\StringTok{ }\DecValTok{0}\NormalTok{) }\OperatorTok{&}\StringTok{ }\NormalTok{(Bj0 }\OperatorTok{==}\StringTok{ }\DecValTok{1}\NormalTok{))}\KeywordTok{.sum}\NormalTok{()     }\CommentTok{# New variable}
\NormalTok{            n00 =}\StringTok{ }\NormalTok{((Ai0 }\OperatorTok{==}\StringTok{ }\DecValTok{0}\NormalTok{) }\OperatorTok{&}\StringTok{ }\NormalTok{(Bj0 }\OperatorTok{==}\StringTok{ }\DecValTok{0}\NormalTok{))}\KeywordTok{.sum}\NormalTok{()     }\CommentTok{# New variable}
\NormalTok{            n0m =}\StringTok{ }\NormalTok{((Ai0 }\OperatorTok{==}\StringTok{ }\DecValTok{0}\NormalTok{) }\OperatorTok{&}\StringTok{ }\NormalTok{(Bj0 }\OperatorTok{==}\StringTok{ }\DecValTok{-1}\NormalTok{))}\KeywordTok{.sum}\NormalTok{()    }\CommentTok{# New variable}
\NormalTok{            nmp =}\StringTok{ }\NormalTok{((Ai0 }\OperatorTok{==}\StringTok{ }\DecValTok{-1}\NormalTok{) }\OperatorTok{&}\StringTok{ }\NormalTok{(Bj0 }\OperatorTok{==}\StringTok{ }\DecValTok{1}\NormalTok{))}\KeywordTok{.sum}\NormalTok{()    }\CommentTok{# New variable}
\NormalTok{            nm0 =}\StringTok{ }\NormalTok{((Ai0 }\OperatorTok{==}\StringTok{ }\DecValTok{-1}\NormalTok{) }\OperatorTok{&}\StringTok{ }\NormalTok{(Bj0 }\OperatorTok{==}\StringTok{ }\DecValTok{0}\NormalTok{))}\KeywordTok{.sum}\NormalTok{()    }\CommentTok{# New variable}
\NormalTok{            nmm =}\StringTok{ }\NormalTok{((Ai0 }\OperatorTok{==}\StringTok{ }\DecValTok{-1}\NormalTok{) }\OperatorTok{&}\StringTok{ }\NormalTok{(Bj0 }\OperatorTok{==}\StringTok{ }\DecValTok{-1}\NormalTok{))}\KeywordTok{.sum}\NormalTok{()   }\CommentTok{# New variable}
\NormalTok{            print npp, np0, npm              }\CommentTok{# Print out extra data}
\NormalTok{            print n0p, n00, n0m              }\CommentTok{# Print out extra data}
\NormalTok{            print nmp, nm0, nmm              }\CommentTok{# Print out extra data}
            \KeywordTok{CHSH.append}\NormalTok{( (Ai}\OperatorTok{*}\NormalTok{Bj)}\KeywordTok{.mean}\NormalTok{())}
            \KeywordTok{QM.append}\NormalTok{( }\OperatorTok{-}\KeywordTok{numpy.cos}\NormalTok{(}\KeywordTok{numpy.radians}\NormalTok{(j}\OperatorTok{-}\NormalTok{i)) )}
\end{Highlighting}
\end{Shaded}

I ran epr-simple, redirecting the output to a text file called
\lq\lq\texttt{data.txt}\rq\rq. In a text editor, I deleted all but 12
lines of that file -- the lines containing the numbers which are read into
R, and then printed out by the R code below. I omit the output tables of
numbers here, to save space.

\begin{Shaded}
\begin{Highlighting}[]
\KeywordTok{setwd}\NormalTok{(}\StringTok{"~/Desktop/Bell/Minkwe/minkwe_v7"}\NormalTok{)}
\NormalTok{data <-}\StringTok{ }\KeywordTok{as.matrix}\NormalTok{(}\KeywordTok{read.table}\NormalTok{(}\StringTok{"data.txt"}\NormalTok{))}
\KeywordTok{colnames}\NormalTok{(data) <-}\StringTok{ }\OtherTok{NULL}
\NormalTok{data[}\DecValTok{1}\OperatorTok{:}\DecValTok{3}\NormalTok{, ]}
\NormalTok{data[}\DecValTok{4}\OperatorTok{:}\DecValTok{6}\NormalTok{, ]}
\NormalTok{data[}\DecValTok{7}\OperatorTok{:}\DecValTok{9}\NormalTok{, ]}
\NormalTok{data[}\DecValTok{10}\OperatorTok{:}\DecValTok{12}\NormalTok{, ]}
\KeywordTok{dim}\NormalTok{(data) <-}\StringTok{ }\KeywordTok{c}\NormalTok{(}\DecValTok{3}\NormalTok{, }\DecValTok{2}\NormalTok{, }\DecValTok{2}\NormalTok{, }\DecValTok{3}\NormalTok{)}
\NormalTok{Outcomes <-}\StringTok{ }\KeywordTok{as.character}\NormalTok{(}\KeywordTok{c}\NormalTok{(}\DecValTok{1}\NormalTok{, }\DecValTok{0}\NormalTok{, }\DecValTok{-1}\NormalTok{))}
\NormalTok{Settings <-}\StringTok{ }\KeywordTok{as.character}\NormalTok{(}\KeywordTok{c}\NormalTok{(}\DecValTok{1}\NormalTok{, }\DecValTok{2}\NormalTok{))}
\NormalTok{dims <-}\StringTok{ }\KeywordTok{list}\NormalTok{(}\DataTypeTok{AliceOut =}\NormalTok{ Outcomes, }\DataTypeTok{AliceIn =}\NormalTok{ Settings, }
             \DataTypeTok{BobIn =}\NormalTok{ Settings, }\DataTypeTok{BobOut =}\NormalTok{ Outcomes)}
\KeywordTok{dimnames}\NormalTok{(data) <-}\StringTok{ }\NormalTok{dims}
\NormalTok{data <-}\StringTok{ }\KeywordTok{aperm}\NormalTok{(data, }\KeywordTok{c}\NormalTok{(}\DecValTok{1}\NormalTok{, }\DecValTok{4}\NormalTok{, }\DecValTok{2}\NormalTok{, }\DecValTok{3}\NormalTok{))}
\NormalTok{data}
\NormalTok{rho <-}\StringTok{ }\ControlFlowTok{function}\NormalTok{(D) (D[}\DecValTok{1}\NormalTok{, }\DecValTok{1}\NormalTok{] }\OperatorTok{+}\StringTok{ }\NormalTok{D[}\DecValTok{3}\NormalTok{, }\DecValTok{3}\NormalTok{] }\OperatorTok{-}\StringTok{ }\NormalTok{D[}\DecValTok{1}\NormalTok{, }\DecValTok{3}\NormalTok{] }\OperatorTok{-}\StringTok{ }\NormalTok{D[}\DecValTok{3}\NormalTok{, }\DecValTok{1}\NormalTok{]) }\OperatorTok{/}
\StringTok{                         }\NormalTok{(D[}\DecValTok{1}\NormalTok{, }\DecValTok{1}\NormalTok{] }\OperatorTok{+}\StringTok{ }\NormalTok{D[}\DecValTok{3}\NormalTok{, }\DecValTok{3}\NormalTok{] }\OperatorTok{+}\StringTok{ }\NormalTok{D[}\DecValTok{1}\NormalTok{, }\DecValTok{3}\NormalTok{] }\OperatorTok{+}\StringTok{ }\NormalTok{D[}\DecValTok{3}\NormalTok{, }\DecValTok{1}\NormalTok{])}
\NormalTok{corrs <-}\StringTok{ }\KeywordTok{matrix}\NormalTok{(}\DecValTok{0}\NormalTok{, }\DecValTok{2}\NormalTok{, }\DecValTok{2}\NormalTok{)}
\ControlFlowTok{for}\NormalTok{(i }\ControlFlowTok{in} \DecValTok{1}\OperatorTok{:}\DecValTok{2}\NormalTok{) \{}\ControlFlowTok{for}\NormalTok{ (j }\ControlFlowTok{in} \DecValTok{1}\OperatorTok{:}\DecValTok{2}\NormalTok{) corrs[i, j] <-}\StringTok{ }\KeywordTok{rho}\NormalTok{(data[ , , i, j])\}}
\NormalTok{contrast <-}\StringTok{ }\KeywordTok{c}\NormalTok{(}\OperatorTok{-}\DecValTok{1}\NormalTok{, }\OperatorTok{+}\DecValTok{1}\NormalTok{, }\DecValTok{-1}\NormalTok{, }\DecValTok{-1}\NormalTok{)}
\NormalTok{S <-}\StringTok{ }\KeywordTok{sum}\NormalTok{(corrs }\OperatorTok{*}\StringTok{ }\NormalTok{contrast)}
\NormalTok{corrs}
\NormalTok{S  }\CommentTok{## observed value of CHSH}
\DecValTok{2} \OperatorTok{*}\StringTok{ }\KeywordTok{sqrt}\NormalTok{(}\DecValTok{2}\NormalTok{)   }\CommentTok{## QM prediction (Tsirelson bound)}
\end{Highlighting}
\end{Shaded}

We see a nice violation of CHSH; $S =2.798796$ . However, a large number of particle
pairs have been rejected.

\begin{Shaded}
\begin{Highlighting}[]
\NormalTok{eta <-}\StringTok{ }\ControlFlowTok{function}\NormalTok{(D) (D[}\DecValTok{1}\NormalTok{, }\DecValTok{1}\NormalTok{] }\OperatorTok{+}\StringTok{ }\NormalTok{D[}\DecValTok{1}\NormalTok{, }\DecValTok{3}\NormalTok{] }\OperatorTok{+}\StringTok{ }\NormalTok{D[}\DecValTok{3}\NormalTok{, }\DecValTok{1}\NormalTok{] }\OperatorTok{+}\StringTok{ }\NormalTok{D[}\DecValTok{3}\NormalTok{, }\DecValTok{3}\NormalTok{])}\OperatorTok{/}\KeywordTok{sum}\NormalTok{(D[ , }\DecValTok{-2}\NormalTok{])}
\NormalTok{etap <-}\StringTok{ }\ControlFlowTok{function}\NormalTok{(D) (D[}\DecValTok{1}\NormalTok{, }\DecValTok{1}\NormalTok{] }\OperatorTok{+}\StringTok{ }\NormalTok{D[}\DecValTok{1}\NormalTok{, }\DecValTok{3}\NormalTok{] }\OperatorTok{+}\StringTok{ }\NormalTok{D[}\DecValTok{3}\NormalTok{, }\DecValTok{1}\NormalTok{] }\OperatorTok{+}\StringTok{ }\NormalTok{D[}\DecValTok{3}\NormalTok{, }\DecValTok{3}\NormalTok{])}\OperatorTok{/}\KeywordTok{sum}\NormalTok{(D[}\OperatorTok{-}\DecValTok{2}\NormalTok{, ])}
\NormalTok{efficiency <-}\StringTok{ }\KeywordTok{matrix}\NormalTok{(}\DecValTok{0}\NormalTok{, }\DecValTok{2}\NormalTok{, }\DecValTok{2}\NormalTok{)}
\ControlFlowTok{for}\NormalTok{(i }\ControlFlowTok{in} \DecValTok{1}\OperatorTok{:}\DecValTok{2}\NormalTok{) \{}\ControlFlowTok{for}\NormalTok{ (j }\ControlFlowTok{in} \DecValTok{1}\OperatorTok{:}\DecValTok{2}\NormalTok{) efficiency[i, j] <-}\StringTok{ }\KeywordTok{eta}\NormalTok{(data[ , , i, j])\}}
\NormalTok{efficiencyp <-}\StringTok{ }\KeywordTok{matrix}\NormalTok{(}\DecValTok{0}\NormalTok{, }\DecValTok{2}\NormalTok{, }\DecValTok{2}\NormalTok{)}
\ControlFlowTok{for}\NormalTok{(i }\ControlFlowTok{in} \DecValTok{1}\OperatorTok{:}\DecValTok{2}\NormalTok{) \{}\ControlFlowTok{for}\NormalTok{ (j }\ControlFlowTok{in} \DecValTok{1}\OperatorTok{:}\DecValTok{2}\NormalTok{) efficiencyp[i, j] <-}\StringTok{ }\KeywordTok{etap}\NormalTok{(data[ , , i, j])\}}
\NormalTok{efficiency; efficiencyp}
\NormalTok{etamin <-}\StringTok{ }\KeywordTok{min}\NormalTok{(efficiency, efficiencyp)}
\NormalTok{etamin}
\end{Highlighting}
\end{Shaded}

It turns out that the minimum over all setting pairs, and over the two
permutations of the set of two parties \{Alice, Bob\}, of the
probability that Party 1 has an outcome given Party 2 has an outcome, is
\(\eta\approx 81\%\).

A \lq\lq\emph{correct}\rq\rq\, bound to the post-selected CHSH quantity
\(S\) is not $2$, but \(4/\eta - 2\) (Larsson, 2014).

\begin{Shaded}
\begin{Highlighting}[]
\NormalTok{S                    }\CommentTok{# CHSH}
\DecValTok{4} \OperatorTok{/}\StringTok{ }\NormalTok{etamin }\OperatorTok{-}\StringTok{ }\DecValTok{2}       \CommentTok{# bound}
\end{Highlighting}
\end{Shaded}

The corrected bound turns out to be about $2.9$, just above the Tsirelson bound
\(2 \sqrt 2 \approx 2.8\). The simulation generates results just below
the bound: at about $2.8$. The observed value of \(S\), the quantum
mechanical prediction \(2\sqrt 2\) and the adjusted CHSH bound are all
quite close together: the simulation model is pretty close to optimal.

\hypertarget{epr-clocked}{%
\section{epr-clocked}\label{epr-clocked}}

The program epr-clocked uses the coincidence loophole (Pascazio, 1986).
Michel Fodje calls this a \lq\lq\emph{clocked experiment}\rq\rq\, where he
means that time is continuous, the times of detection of particles are
random and unpredictable. (I would have preferred to reserve the word
\lq\lq\emph{clocked}\rq\rq\, as synonym for \lq\lq\emph{pulsed}\rq\rq).
Because Alice and Bob's particles have different, random, delays
(influenced by the detector settings which they meet), one cannot
identify which particles were originally part of which particle pairs.
Moreover, a small number of particles did not get detected at all,
compounding this problem.

The experimenter scans through the data looking for detections which are
within some short time interval of one another. This is called the
detection window. Unpaired detections are discarded.

I ran the program, setting the spin to equal 0.5, and an experiment of
duration 10 seconds. I let Alice and Bob use the settings for a CHSH
experiment: Alice uses angles 0 and 90 degrees, Bob uses angles 45 and
135 degrees. (As in epr-simple, Michel Fodje took the angles
corresponding to a polarization experiment instead of a spin
experiment). I set the numpy random seed to the values \lq\lq1234\rq\rq,
\lq\lq2345\rq\rq, and \lq\lq3456\rq\rq prior to running the source
program, Alice's station program, and Bob's station programme
respectively. This should make my results exactly reproducible
\ldots\ but it doesn't quite achieve that, because the 10 second duration of the
experiment is ten seconds in \lq\lq\emph{real
time}\rq\rq. It therefore depends on queries by the program of the
actual time in the real world outside the computer, and this process
itself can take different lengths of time on each new run. However, the
difference between the data obtained in different runs (with the same
seed) should be negligeable: the total number of particle pairs will
vary slightly, but their initial segments should coincide.

In order to get a strictly reproducible simulation, I rewrote one
section of the program \lq\lq\texttt{source.py}\rq\rq. Below is my
replacement code. Instead of running for 10 seconds of real time, the
code simply generates exactly 200 000 emissions.

\begin{Shaded}
\begin{Highlighting}[]
\NormalTok{    def }\KeywordTok{run}\NormalTok{(self, }\DataTypeTok{duration=}\FloatTok{10.0}\NormalTok{)}\OperatorTok{:}
\StringTok{        }\NormalTok{N =}\StringTok{ }\DecValTok{200000}
\NormalTok{        n =}\StringTok{ }\DecValTok{1}
\NormalTok{        print }\StringTok{"Generating spin-%0.1f particle pairs"}\NormalTok{ % (self.n}\OperatorTok{/}\FloatTok{2.0}\NormalTok{)}
        \ControlFlowTok{while}\NormalTok{ n }\OperatorTok{<=}\StringTok{ }\NormalTok{N}\OperatorTok{:}
\StringTok{            }\KeywordTok{self.emit}\NormalTok{()}
\NormalTok{            n =}\StringTok{ }\NormalTok{n }\OperatorTok{+}\StringTok{ }\DecValTok{1}
        \KeywordTok{self.save}\NormalTok{(}\StringTok{'SrcLeft.npy.gz'}\NormalTok{, }\KeywordTok{numpy.array}\NormalTok{(self.left))}
        \KeywordTok{self.save}\NormalTok{(}\StringTok{'SrcRight.npy.gz'}\NormalTok{, }\KeywordTok{numpy.array}\NormalTok{(self.right))}
\NormalTok{        print}
\NormalTok{        print }\StringTok{"%d particles in 'SrcLeft.npy.gz'"}\NormalTok{ % (}\KeywordTok{len}\NormalTok{(self.left))}
\NormalTok{        print }\StringTok{"%d particles in 'SrcRight.npy.gz'"}\NormalTok{ % (}\KeywordTok{len}\NormalTok{(self.right))}
\end{Highlighting}
\end{Shaded}

The standard output gave me the following information:

\begin{Shaded}
\begin{Highlighting}[]
\CommentTok{# No. of detected particles}
\CommentTok{#   Alice:          199994}
\CommentTok{#     Bob:          199993}
\CommentTok{# Calculation of expectation values}
\CommentTok{# Settings       N_ab}
\CommentTok{#    0,  45      27416}
\CommentTok{#    0, 135      27512}
\CommentTok{#   90,  45      27345}
\CommentTok{#   90, 135      27425}
\CommentTok{#   CHSH: <= 2.0, Sim: 2.790, QM: 2.828}
\end{Highlighting}
\end{Shaded}

Notice the total number of detected particles on either side, and the
total numbers of coincidences for each of the four setting pairs. The
total number of coincidence pairs is a bit more than 100 thousand; the
total number of detections on either side is almost 200 thousand. We
have a rather poor experimental efficiency of about 55\%.

\begin{Shaded}
\begin{Highlighting}[]
\NormalTok{Npairs <-}\StringTok{ }\DecValTok{27416} \OperatorTok{+}\StringTok{ }\DecValTok{27512} \OperatorTok{+}\StringTok{ }\DecValTok{27345} \OperatorTok{+}\StringTok{ }\DecValTok{27425}
\NormalTok{Nsingles <-}\StringTok{ }\DecValTok{199994}
\NormalTok{Npairs; Nsingles}
\NormalTok{gamma <-}\StringTok{ }\NormalTok{Npairs}\OperatorTok{/}\NormalTok{Nsingles}
\NormalTok{gamma}
\DecValTok{6} \OperatorTok{/}\StringTok{ }\NormalTok{gamma }\OperatorTok{-}\StringTok{ }\DecValTok{4}
\end{Highlighting}
\end{Shaded}

The \lq\lq\emph{correct}\rq\rq\, bound to the coincidence-selected CHSH
quantity \(S\) is not 2. In fact, we do not know it exactly, but a
correct bound is conjectured to be \(6/\gamma - 4\) (Larsson, 2014).
Here, \(\gamma\) is the effective efficiency of the experiment measured
as the chance that a detected particle on one side of the experiment
will be accepted as part of a coincidence pair. Notice that at
\(\gamma = 1\) (full efficiency) the adjusted bound is equal to the
usual bound 2, but that as it decreases from 100\% the bound rapidly
increases. At \(\gamma = 3/4\) it reaches its natural maximum of 4.

At the observed efficiency of about 55\%, the corrected CHSH bound in
this experiment is close to 7, far above the natural and absolute bound
4.

Because the efficiency is lower than in \texttt{epr-simple}, while the
proper bound (adjusted CHSH) is higher, this experiment is a good deal
worse than the previous one in terms of efficiency. It wouldn't be too
difficult, at this level of experimental efficiency, to tune parameters
of this model so as to get the observed value of CHSH up to its natural
maximum of 4.

Incidentally, running epr-clocked many times, I experienced quite a few
failures of the program \lq\lq\texttt{analyse.py}\rq\rq which is
supposed to extract the coincidence pairs from the two data files. It
seems that the Larsson algorithm for finding the pairs, which Fodje has
adopted for this part of the data analysis, is failing in some
circumstances. I could not find out what was the cause of this.

Fortunately it is now rather easy to import \texttt{Numpy}
(\lq\lq\emph{numerical python}\rq\rq) binary data files into R using the
package \lq\lq\texttt{RcppCNPy}\rq\rq. It should also not be difficult
to find a suitable alternative to Larsson's matching algorithm in the
computer science literature and probably freely available in C++
libraries. Algorithms written in C++ can often easily be made available
in R via \lq\lq\texttt{Rcpp}\rq\rq. Hence one could replace Michel
Fodje's \lq\lq\texttt{analyse.py}\rq\rq\, by one's own data analysis
script; this would also allow a \lq\lq\emph{proper}\rq\rq computation of
the efficiency \(\gamma\), taking the minimum over the efficiencies for
each setting pair and both permutations of the two parties Alice and
Bob.

\hypertarget{references}{%
\section{References}\label{references}}

R.D. Gill (2015, 2020), Pearle's Hidden-Variable Model Revisited.
\emph{Entropy} 2020, {\bf 22} (1), 1,
\url{ https://doi.org/10.3390/e22010001}; 
\url{http://arxiv.org/abs/1505.04431}

J.-A. Larsson (2014), Loopholes in Bell Inequality Tests of Local
Realism, \emph{J. Phys. A} {\bf 47} 424003 (2014).
\url{http://arxiv.org/abs/1407.0363}

S. Pascazio (1986), Time and Bell-type inequalities, \emph{Phys. Letters
A} \textbf{118}, 47-53.

P.M. Pearle (1970), Hidden-Variable Example Based upon Data Rejection,
\emph{Phys. Rev.~D} \textbf{2}, 1418-1425.

\hypertarget{appendix}{%
\section{Appendix}\label{appendix}}

The appendices below contain mathematical formulas of the two simulation
models. I have done my best to extract these faithfully from the
original Python code and accompanying explanations by the author Michel
Fodje. I have earlier published translations into the R language:
\url{http://rpubs.com/gill1109/epr-simple},
\url{http://rpubs.com/gill1109/epr-clocked-core},
\url{http://rpubs.com/gill1109/epr-clocked-full}

\hypertarget{appendix-epr-simple}{%
\section{Appendix: epr-simple}\label{appendix-epr-simple}}

Here is a little simulation experiment with (my interpretation in R of)
\texttt{epr-simple}. I plot the simulated correlation function and also
the acceptance rate. With an effective sample size of
\(N \approx 0.8 \times 10^6\), the statistical error in simulated
estimated correlation coefficients is roughly of size 0.001, well below
the resolution of the graphs plotted below. Thus the small visible
deviations from the theoretical negative cosine curve are for real. I
simplify the model by taking spin = 1/2. Formulas are further simplified
by a sign flip of all measurement outcomes, which by the symmetries of
the model, does not change the observed data statistics.

\begin{Shaded}
\begin{Highlighting}[]
\KeywordTok{set.seed}\NormalTok{(}\DecValTok{9875}\NormalTok{)}
\CommentTok{## For reproducibility. Replace integer seed by your own, }
\CommentTok{## or delete this line and let your computer dream up one for you }
\CommentTok{## (it will use system time + process ID).}
\CommentTok{## Measurement angles for setting 'a': }
\CommentTok{##        directions in the equatorial plane}
\NormalTok{angles <-}\StringTok{ }\KeywordTok{seq}\NormalTok{(}\DataTypeTok{from =} \DecValTok{0}\NormalTok{, }\DataTypeTok{to =} \DecValTok{360}\NormalTok{, }\DataTypeTok{by =} \DecValTok{1}\NormalTok{) }\OperatorTok{*}\StringTok{ }\DecValTok{2} \OperatorTok{*}\StringTok{ }\NormalTok{pi}\OperatorTok{/}\DecValTok{360}
\NormalTok{K <-}\StringTok{ }\KeywordTok{length}\NormalTok{(angles)}
\NormalTok{corrs <-}\StringTok{ }\KeywordTok{numeric}\NormalTok{(K)  }\CommentTok{## Container for correlations}
\NormalTok{Ns <-}\StringTok{ }\KeywordTok{numeric}\NormalTok{(K)     }\CommentTok{## Container for number of states}
\NormalTok{beta <-}\StringTok{ }\DecValTok{0} \OperatorTok{*}\StringTok{ }\DecValTok{2} \OperatorTok{*}\StringTok{ }\NormalTok{pi}\OperatorTok{/}\DecValTok{360}  \CommentTok{## Measurement direction 'b' fixed, }
\CommentTok{##                         in equatorial plane}
\NormalTok{b <-}\StringTok{ }\KeywordTok{c}\NormalTok{(}\KeywordTok{cos}\NormalTok{(beta), }\KeywordTok{sin}\NormalTok{(beta))  }\CommentTok{## Measurement vector 'b'}
\NormalTok{M <-}\StringTok{ }\DecValTok{10}\OperatorTok{^}\DecValTok{6}  \CommentTok{## Size of "pre-ensemble"}
\CommentTok{## Use the same, single sample of 'M' realizations of hidden}
\CommentTok{## states for all measurement directions. This saves a lot of time,}
\CommentTok{## and reduces variance when we look at *differences*.}
\NormalTok{e <-}\StringTok{ }\KeywordTok{runif}\NormalTok{(M, }\DecValTok{0}\NormalTok{, }\DecValTok{2}\OperatorTok{*}\NormalTok{pi)}
\NormalTok{ep <-}\StringTok{ }\NormalTok{e }\OperatorTok{+}\StringTok{ }\NormalTok{pi}
\NormalTok{U <-}\StringTok{ }\KeywordTok{runif}\NormalTok{(M)}
\NormalTok{p <-}\StringTok{ }\NormalTok{(}\KeywordTok{sin}\NormalTok{(U }\OperatorTok{*}\StringTok{ }\NormalTok{pi }\OperatorTok{/}\StringTok{ }\DecValTok{2}\NormalTok{)}\OperatorTok{^}\DecValTok{2}\NormalTok{)}\OperatorTok{/}\DecValTok{2}
\CommentTok{## Loop through measurement vectors 'a' }
\CommentTok{##         (except last = 360 degrees = first)}
\ControlFlowTok{for}\NormalTok{ (i }\ControlFlowTok{in} \DecValTok{1}\OperatorTok{:}\NormalTok{(K }\OperatorTok{-}\StringTok{ }\DecValTok{1}\NormalTok{)) \{}
\NormalTok{  alpha <-}\StringTok{ }\NormalTok{angles[i]}
\NormalTok{  ca <-}\StringTok{ }\KeywordTok{cos}\NormalTok{(alpha }\OperatorTok{-}\StringTok{ }\NormalTok{e)}
\NormalTok{  cb <-}\StringTok{ }\KeywordTok{cos}\NormalTok{(beta }\OperatorTok{-}\StringTok{ }\NormalTok{ep)}
\NormalTok{  A <-}\StringTok{ }\KeywordTok{ifelse}\NormalTok{(}\KeywordTok{abs}\NormalTok{(ca) }\OperatorTok{>}\StringTok{ }\NormalTok{p, }\KeywordTok{sign}\NormalTok{(ca), }\DecValTok{0}\NormalTok{)}
\NormalTok{  B <-}\StringTok{ }\KeywordTok{ifelse}\NormalTok{(}\KeywordTok{abs}\NormalTok{(cb) }\OperatorTok{>}\StringTok{ }\NormalTok{p, }\KeywordTok{sign}\NormalTok{(cb), }\DecValTok{0}\NormalTok{)}
\NormalTok{  AB <-}\StringTok{ }\NormalTok{A}\OperatorTok{*}\NormalTok{B}
\NormalTok{  good <-}\StringTok{ }\NormalTok{AB }\OperatorTok{!=}\StringTok{ }\DecValTok{0}
\NormalTok{  corrs[i] <-}\StringTok{ }\KeywordTok{mean}\NormalTok{(AB[good])}
\NormalTok{  Ns[i] <-}\StringTok{ }\KeywordTok{sum}\NormalTok{(good)}
\NormalTok{\}}
\NormalTok{corrs[K] <-}\StringTok{ }\NormalTok{corrs[}\DecValTok{1}\NormalTok{]}
\NormalTok{Ns[K] <-}\StringTok{ }\NormalTok{Ns[}\DecValTok{1}\NormalTok{]}
\KeywordTok{plot}\NormalTok{(angles }\OperatorTok{*}\StringTok{ }\DecValTok{180}\OperatorTok{/}\NormalTok{pi, corrs, }\DataTypeTok{type =} \StringTok{"l"}\NormalTok{, }\DataTypeTok{col =} \StringTok{"blue"}\NormalTok{, }
     \DataTypeTok{xlab =} \StringTok{"Angle (degrees)"}\NormalTok{, }\DataTypeTok{ylab =} \StringTok{"Correlation"}\NormalTok{)}
\KeywordTok{lines}\NormalTok{(angles }\OperatorTok{*}\StringTok{ }\DecValTok{180}\OperatorTok{/}\NormalTok{pi, }\OperatorTok{-}\StringTok{ }\KeywordTok{cos}\NormalTok{(angles), }\DataTypeTok{col =} \StringTok{"black"}\NormalTok{)}
\KeywordTok{legend}\NormalTok{(}\DataTypeTok{x =} \DecValTok{0}\NormalTok{, }\DataTypeTok{y =} \FloatTok{1.0}\NormalTok{, }\DataTypeTok{legend =} \KeywordTok{c}\NormalTok{(}\StringTok{"epr-simple"}\NormalTok{, }\StringTok{"neg cosine"}\NormalTok{), }
       \DataTypeTok{text.col =} \KeywordTok{c}\NormalTok{(}\StringTok{"blue"}\NormalTok{, }\StringTok{"black"}\NormalTok{), }\DataTypeTok{lty =} \DecValTok{1}\NormalTok{, }\DataTypeTok{col =} \KeywordTok{c}\NormalTok{(}\StringTok{"blue"}\NormalTok{, }\StringTok{"black"}\NormalTok{))}
\KeywordTok{plot}\NormalTok{(angles }\OperatorTok{*}\StringTok{ }\DecValTok{180}\OperatorTok{/}\NormalTok{pi, Ns }\OperatorTok{/}\StringTok{ }\NormalTok{M, }\DataTypeTok{type =} \StringTok{"l"}\NormalTok{, }\DataTypeTok{col =} \StringTok{"blue"}\NormalTok{, }
     \DataTypeTok{xlab =} \StringTok{"Angle (degrees)"}\NormalTok{,} \DataTypeTok{ylim =} \KeywordTok{c}\NormalTok{(}\DecValTok{0}\NormalTok{, }\DecValTok{1}\NormalTok{))}
\end{Highlighting}
\end{Shaded}
\begin{figure}[H]
    \centering
    \begin{minipage}{0.5\textwidth}
        \centering
        \includegraphics[width=0.95\textwidth]{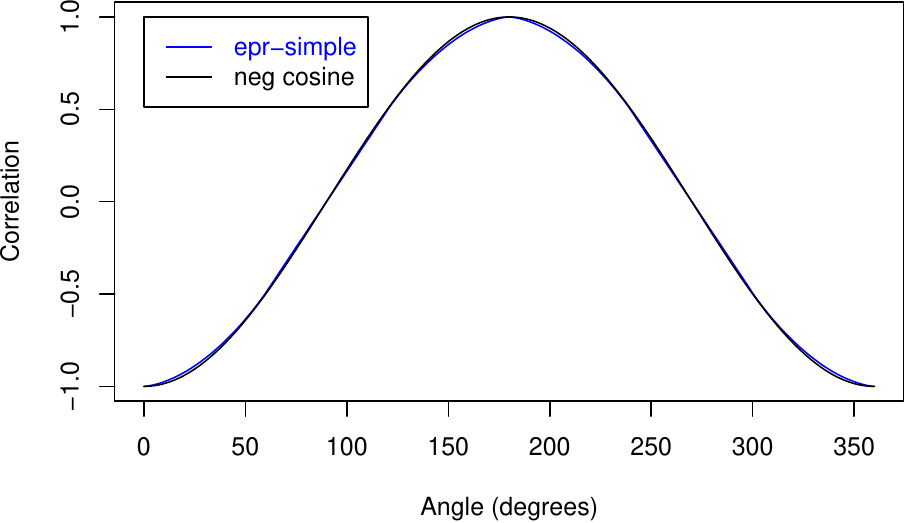} 
        \caption{epr-simple}
    \end{minipage}\hfill
    \begin{minipage}{0.5\textwidth}
        \centering
        \includegraphics[width=0.95\textwidth]{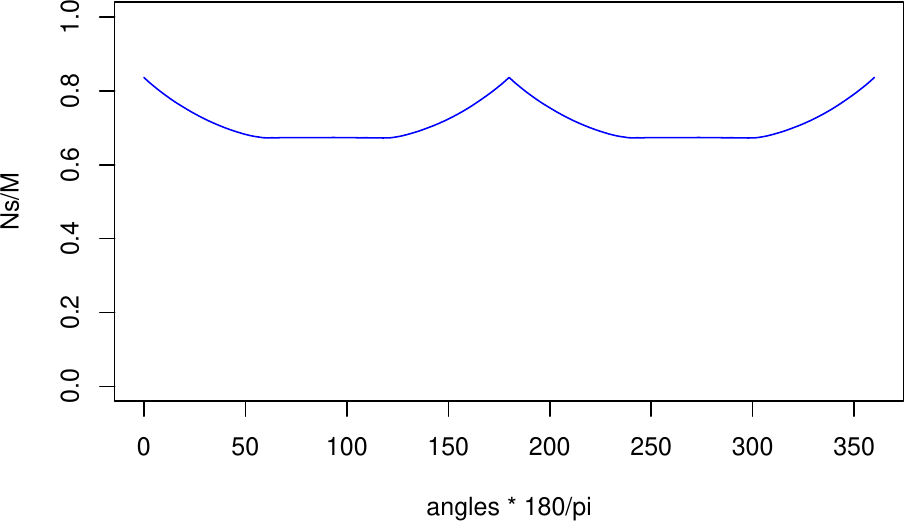} 
        \caption{Rate of detected particle pairs}
    \end{minipage}
\end{figure}

The detection loophole model used here is very simple. There is a hidden
variable \(E\) uniformly distributed in \([0, 2\pi]\). Independently
thereof, there is a second hidden variable \(P\) taking values in
\([0, 1/2]\). Its distribution is determined by the relation
\(P = \sin^2((\pi/2) U)/2\) where \(U\) is uniform on \([0, 1]\). Alice
and Bob's measurement outcomes are \(\text{sign}\cos(E-\alpha)\) and
\(\text{sign}\cos(E-\beta+ \pi)\) respectively, \emph{if} each of their
particles is detected. Alice's particle is detected if and only if
\(\text{abs}(\cos(E-\alpha)) > P\) and Bob's if and only if
\(\text{abs}(\cos(E-\beta)) > P\).

Pearle (1970) characterized mathematically the set of all probability
distributions of \(P\) which would give us the singlet correlations
\emph{exactly} (and for measurement directions in space, not just in the
plane). He also picks out one particularly simple model in the class.
His special choice has \(P = (2/\sqrt V) -1 \in [0, 1]\) where \(V\) is
uniform on \([1, 4]\), first expressed in this way by myself in 2014,
see \url{http://rpubs.com/gill1109/pearle} and Gill (2015.)

Below, the simulation is modified accordingly: just one line of code is
altered. Now the experimental and theoretical curves are
indistinguishable.

\begin{Shaded}
\begin{Highlighting}[]
\CommentTok{# p <- (sin(U * pi / 2)^2)/2  ## epr-simple}
\NormalTok{p <-}\StringTok{ }\DecValTok{2}\OperatorTok{/}\KeywordTok{sqrt}\NormalTok{(}\DecValTok{1} \OperatorTok{+}\StringTok{ }\DecValTok{3} \OperatorTok{*}\StringTok{ }\NormalTok{U) }\OperatorTok{-}\StringTok{ }\DecValTok{1}    \CommentTok{## Pearle (1970) model}
\CommentTok{## Loop through measurement vectors 'a' }
\CommentTok{##           (except last = 360 degrees = first)}
\ControlFlowTok{for}\NormalTok{ (i }\ControlFlowTok{in} \DecValTok{1}\OperatorTok{:}\NormalTok{(K }\OperatorTok{-}\StringTok{ }\DecValTok{1}\NormalTok{)) \{}
\NormalTok{  alpha <-}\StringTok{ }\NormalTok{angles[i]}
\NormalTok{  ca <-}\StringTok{ }\KeywordTok{cos}\NormalTok{(alpha }\OperatorTok{-}\StringTok{ }\NormalTok{e)}
\NormalTok{  cb <-}\StringTok{ }\KeywordTok{cos}\NormalTok{(beta }\OperatorTok{-}\StringTok{ }\NormalTok{ep)}
\NormalTok{  A <-}\StringTok{ }\KeywordTok{ifelse}\NormalTok{(}\KeywordTok{abs}\NormalTok{(ca) }\OperatorTok{>}\StringTok{ }\NormalTok{p, }\KeywordTok{sign}\NormalTok{(ca), }\DecValTok{0}\NormalTok{)}
\NormalTok{  B <-}\StringTok{ }\KeywordTok{ifelse}\NormalTok{(}\KeywordTok{abs}\NormalTok{(cb) }\OperatorTok{>}\StringTok{ }\NormalTok{p, }\KeywordTok{sign}\NormalTok{(cb), }\DecValTok{0}\NormalTok{)}
\NormalTok{  AB <-}\StringTok{ }\NormalTok{A}\OperatorTok{*}\NormalTok{B}
\NormalTok{  good <-}\StringTok{ }\NormalTok{AB }\OperatorTok{!=}\StringTok{ }\DecValTok{0}
\NormalTok{  corrs[i] <-}\StringTok{ }\KeywordTok{mean}\NormalTok{(AB[good])}
\NormalTok{  Ns[i] <-}\StringTok{ }\KeywordTok{sum}\NormalTok{(good)}
\NormalTok{\}}
\NormalTok{corrs[K] <-}\StringTok{ }\NormalTok{corrs[}\DecValTok{1}\NormalTok{]}
\NormalTok{Ns[K] <-}\StringTok{ }\NormalTok{Ns[}\DecValTok{1}\NormalTok{]}
\KeywordTok{plot}\NormalTok{(angles }\OperatorTok{*}\StringTok{ }\DecValTok{180}\OperatorTok{/}\NormalTok{pi, corrs, }\DataTypeTok{type =} \StringTok{"l"}\NormalTok{, }\DataTypeTok{col =} \StringTok{"blue"}\NormalTok{, }
     \DataTypeTok{xlab =} \StringTok{"Angle (degrees)"}\NormalTok{, }\DataTypeTok{ylab =} \StringTok{"Correlation"}\NormalTok{)}
\KeywordTok{lines}\NormalTok{(angles }\OperatorTok{*}\StringTok{ }\DecValTok{180}\OperatorTok{/}\NormalTok{pi, }\OperatorTok{-}\StringTok{ }\KeywordTok{cos}\NormalTok{(angles), }\DataTypeTok{col =} \StringTok{"black"}\NormalTok{)}
\KeywordTok{legend}\NormalTok{(}\DataTypeTok{x =} \DecValTok{0}\NormalTok{, }\DataTypeTok{y =} \FloatTok{1.0}\NormalTok{, }\DataTypeTok{legend =} \KeywordTok{c}\NormalTok{(}\StringTok{"Pearle"}\NormalTok{, }\StringTok{"neg cosine"}\NormalTok{), }
       \DataTypeTok{text.col =} \KeywordTok{c}\NormalTok{(}\StringTok{"blue"}\NormalTok{, }\StringTok{"black"}\NormalTok{), }\DataTypeTok{lty =} \DecValTok{1}\NormalTok{, }\DataTypeTok{col =} \KeywordTok{c}\NormalTok{(}\StringTok{"blue"}\NormalTok{, }\StringTok{"black"}\NormalTok{))}
\KeywordTok{plot}\NormalTok{(angles }\OperatorTok{*}\StringTok{ }\DecValTok{180}\OperatorTok{/}\NormalTok{pi, Ns }\OperatorTok{/}\StringTok{ }\NormalTok{M, }\DataTypeTok{type =} \StringTok{"l"}\NormalTok{, }\DataTypeTok{col =} \StringTok{"blue"}\NormalTok{, }
     \DataTypeTok{xlab =} \StringTok{"Angle (degrees)"}\NormalTok{,}
     \DataTypeTok{main =} \StringTok{"Rate of detected particle pairs"}\NormalTok{, }\DataTypeTok{ylim =} \KeywordTok{c}\NormalTok{(}\DecValTok{0}\NormalTok{, }\DecValTok{1}\NormalTok{))}
\end{Highlighting}
\end{Shaded}
\begin{figure}[H]
    \centering
    \begin{minipage}{0.5\textwidth}
        \centering
        \includegraphics[width=0.95\textwidth]{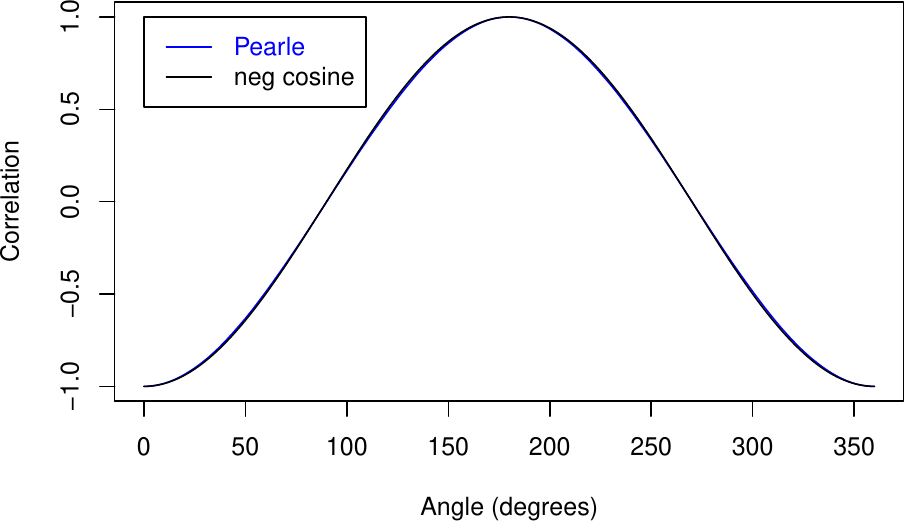} 
        \caption{Pearle}
    \end{minipage}\hfill
    \begin{minipage}{0.5\textwidth}
        \centering
        \includegraphics[width=0.95\textwidth]{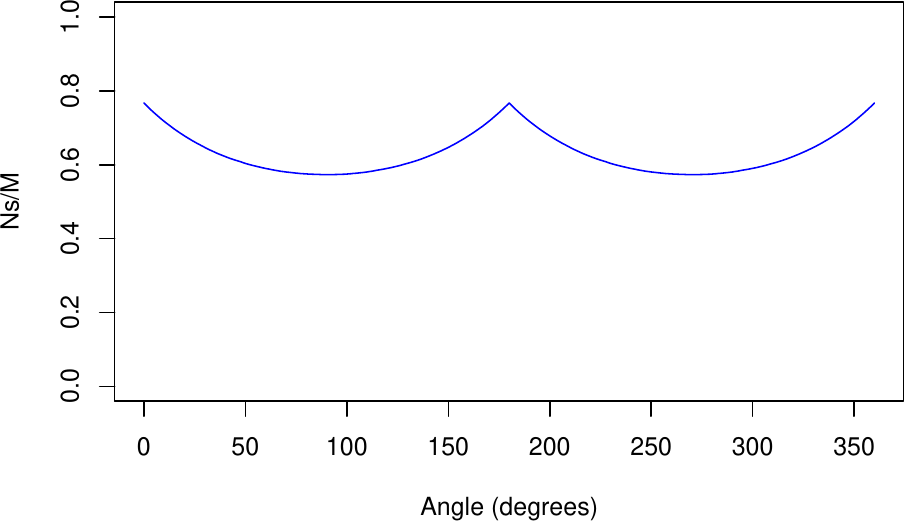} 
        \caption{Rate of detected particle pairs}
    \end{minipage}
\end{figure}

\section{Appendix: epr-clocked}\label{appendix-epr-clocked}

Here is the core part of (my interpretation in R of) epr-clocked: the
emission of a pair of particles, supposing that both are detected and
identified as belonging to a pair. However, if the detection times of
the two particles are too far apart (larger than the so-called
coincidence window), the pair is rejected. This would correspond to
running the full \texttt{epr-clocked}\, model with very low emission
rate. I have simplified the model by fixing spin = 1/2.
To save computer time and memory, I re-use the hidden variables \(E\)
(\lq\lq\texttt{e}\rq\rq) and \(U\) from \texttt{epr-simple}.

Again, I plot the simulated correlation function and also the acceptance
rate.

\begin{Shaded}
\begin{Highlighting}[]
\NormalTok{coincWindow <-}\StringTok{ }\FloatTok{0.0004}
\NormalTok{ts <-}\StringTok{ }\NormalTok{pi }\OperatorTok{*}\StringTok{ }\FloatTok{0.03}  \CommentTok{## timescale}
\NormalTok{asym <-}\StringTok{ }\FloatTok{0.98}     \CommentTok{## asymmetry parameter}
\NormalTok{p <-}\StringTok{ }\FloatTok{0.5} \OperatorTok{*}\StringTok{ }\KeywordTok{sin}\NormalTok{(U }\OperatorTok{*}\StringTok{ }\NormalTok{pi }\OperatorTok{/}\StringTok{ }\DecValTok{6}\NormalTok{)}\OperatorTok{^}\DecValTok{2}
\NormalTok{ml <-}\StringTok{ }\KeywordTok{runif}\NormalTok{(M, asym, }\DecValTok{1}\NormalTok{)     }\CommentTok{## small random jitter, left}
\NormalTok{mr <-}\StringTok{ }\KeywordTok{runif}\NormalTok{(M, asym, }\DecValTok{1}\NormalTok{)     }\CommentTok{## small random jitter, right}
\CommentTok{## Loop through measurement vectors 'a' }
\CommentTok{##      (except last = 360 degrees = first)}
\ControlFlowTok{for}\NormalTok{ (i }\ControlFlowTok{in} \DecValTok{1}\OperatorTok{:}\NormalTok{(K }\OperatorTok{-}\StringTok{ }\DecValTok{1}\NormalTok{)) \{}
\NormalTok{    alpha <-}\StringTok{ }\NormalTok{angles[i]}
\NormalTok{    Cl <-}\StringTok{ }\NormalTok{(}\FloatTok{0.5}\OperatorTok{/}\NormalTok{pi) }\OperatorTok{*}\StringTok{ }\KeywordTok{cos}\NormalTok{(e }\OperatorTok{-}\StringTok{ }\NormalTok{alpha)  }\CommentTok{## cos(e-a), left}
\NormalTok{    Cr <-}\StringTok{ }\NormalTok{(}\FloatTok{0.5}\OperatorTok{/}\NormalTok{pi) }\OperatorTok{*}\StringTok{ }\KeywordTok{cos}\NormalTok{(ep }\OperatorTok{-}\StringTok{ }\NormalTok{beta)   }\CommentTok{## cos(e-a), right}
\NormalTok{    tdl <-}\StringTok{ }\NormalTok{ts }\OperatorTok{*}\StringTok{ }\KeywordTok{pmax}\NormalTok{(ml }\OperatorTok{*}\StringTok{ }\NormalTok{p }\OperatorTok{-}\StringTok{ }\KeywordTok{abs}\NormalTok{(Cl), }\DecValTok{0}\NormalTok{)  }\CommentTok{## time delays, left}
\NormalTok{    tdr <-}\StringTok{ }\NormalTok{ts }\OperatorTok{*}\StringTok{ }\KeywordTok{pmax}\NormalTok{(mr }\OperatorTok{*}\StringTok{ }\NormalTok{p }\OperatorTok{-}\StringTok{ }\KeywordTok{abs}\NormalTok{(Cr), }\DecValTok{0}\NormalTok{)  }\CommentTok{## time delays, right}
\NormalTok{    A <-}\StringTok{ }\KeywordTok{sign}\NormalTok{(Cl)               }\CommentTok{## measurement outcomes, left}
\NormalTok{    B <-}\StringTok{ }\KeywordTok{sign}\NormalTok{(Cr)               }\CommentTok{## measurement outcomes, right}
\NormalTok{    AB <-}\StringTok{ }\NormalTok{A }\OperatorTok{*}\StringTok{ }\NormalTok{B                 }\CommentTok{## product of outcomes}
\NormalTok{    good <-}\StringTok{ }\KeywordTok{abs}\NormalTok{(tdl}\OperatorTok{-}\NormalTok{tdr) }\OperatorTok{<}\StringTok{ }\NormalTok{coincWindow}
\NormalTok{    corrs[i] <-}\StringTok{ }\KeywordTok{mean}\NormalTok{(AB[good])}
\NormalTok{    Ns[i] <-}\StringTok{ }\KeywordTok{sum}\NormalTok{(good)}
\NormalTok{\}}
\NormalTok{corrs[K] <-}\StringTok{ }\NormalTok{corrs[}\DecValTok{1}\NormalTok{]}
\NormalTok{Ns[K] <-}\StringTok{ }\NormalTok{Ns[}\DecValTok{1}\NormalTok{]}
\KeywordTok{plot}\NormalTok{(angles }\OperatorTok{*}\StringTok{ }\DecValTok{180}\OperatorTok{/}\NormalTok{pi, corrs, }\DataTypeTok{type =} \StringTok{"l"}\NormalTok{, }\DataTypeTok{col =} \StringTok{"blue"}\NormalTok{, }
     \DataTypeTok{main =} \StringTok{"epr-clocked"}\NormalTok{, }
     \DataTypeTok{xlab =} \StringTok{"Angle (degrees)"}\NormalTok{, }\DataTypeTok{ylab =} \StringTok{"Correlation"}\NormalTok{)}
\KeywordTok{lines}\NormalTok{(angles }\OperatorTok{*}\StringTok{ }\DecValTok{180}\OperatorTok{/}\NormalTok{pi, }\OperatorTok{-}\StringTok{ }\KeywordTok{cos}\NormalTok{(angles), }\DataTypeTok{col =} \StringTok{"black"}\NormalTok{)}
\KeywordTok{legend}\NormalTok{(}\DataTypeTok{x =} \DecValTok{0}\NormalTok{, }\DataTypeTok{y =} \FloatTok{1.0}\NormalTok{, }\DataTypeTok{legend =} \KeywordTok{c}\NormalTok{(}\StringTok{"epr-clocked"}\NormalTok{, }\StringTok{"neg cosine"}\NormalTok{), }
       \DataTypeTok{text.col =} \KeywordTok{c}\NormalTok{(}\StringTok{"blue"}\NormalTok{, }\StringTok{"black"}\NormalTok{), }\DataTypeTok{lty =} \DecValTok{1}\NormalTok{, }\DataTypeTok{col =} \KeywordTok{c}\NormalTok{(}\StringTok{"blue"}\NormalTok{, }\StringTok{"black"}\NormalTok{))}
\KeywordTok{plot}\NormalTok{(angles }\OperatorTok{*}\StringTok{ }\DecValTok{180}\OperatorTok{/}\NormalTok{pi, Ns }\OperatorTok{/}\StringTok{ }\NormalTok{M, }\DataTypeTok{type =} \StringTok{"l"}\NormalTok{, }\DataTypeTok{col =} \StringTok{"blue"}\NormalTok{, }
     \DataTypeTok{xlab =} \StringTok{"Angle (degrees)"}\NormalTok{,}
     \DataTypeTok{main =} \StringTok{"Rate of accepted particle pairs"}\NormalTok{, }\DataTypeTok{ylim =} \KeywordTok{c}\NormalTok{(}\DecValTok{0}\NormalTok{, }\DecValTok{1}\NormalTok{))}
\end{Highlighting}
\end{Shaded}
\begin{figure}[H]
    \centering
    \begin{minipage}{0.5\textwidth}
        \centering
        \includegraphics[width=0.95\textwidth]{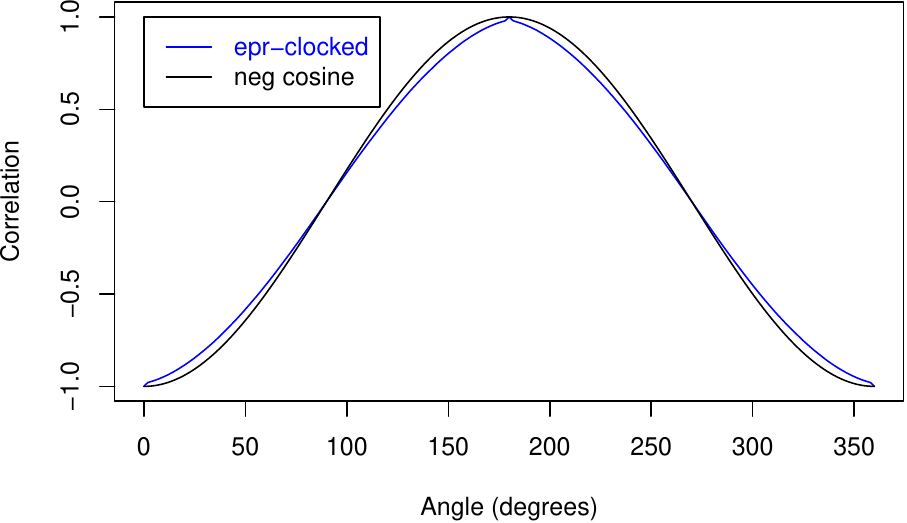} 
        \caption{epr-clocked}
    \end{minipage}\hfill
    \begin{minipage}{0.5\textwidth}
        \centering
        \includegraphics[width=0.95\textwidth]{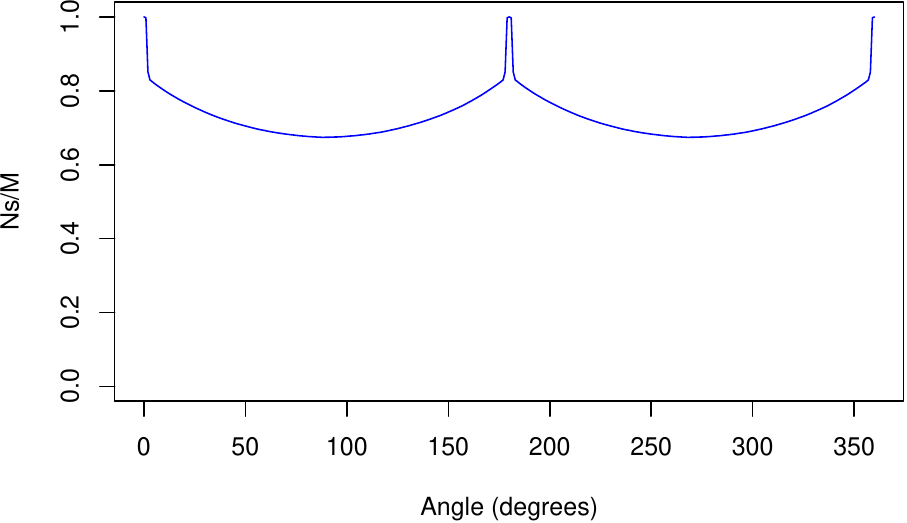} 
        \caption{Rate of detected particle pairs}
    \end{minipage}
\end{figure}

I now remove the \lq\lq\emph{small random jitter}\rq\rq, effectively
setting the \lq\lq\emph{asymmetry parameter}\rq\rq\, to 1. I rescale time
so that the model is finally described in terms of a couple of standard
probability distributions and two arbitrary constants.

\begin{Shaded}
\begin{Highlighting}[]
\NormalTok{coincWindow <-}\StringTok{ }\FloatTok{0.034}
\NormalTok{p <-}\StringTok{ }\DecValTok{8} \OperatorTok{*}\StringTok{ }\NormalTok{p  }\CommentTok{## Now p lies in the interval [0, 1]}
\CommentTok{## Loop through measurement vectors 'a' }
\CommentTok{##            (except last = 360 degrees = first)}
\ControlFlowTok{for}\NormalTok{ (i }\ControlFlowTok{in} \DecValTok{1}\OperatorTok{:}\NormalTok{(K }\OperatorTok{-}\StringTok{ }\DecValTok{1}\NormalTok{)) \{}
\NormalTok{    alpha <-}\StringTok{ }\NormalTok{angles[i]}
\NormalTok{    Cl <-}\StringTok{ }\KeywordTok{cos}\NormalTok{(e }\OperatorTok{-}\StringTok{ }\NormalTok{alpha)    }\CommentTok{## cos(e-a), left}
\NormalTok{    Cr <-}\StringTok{ }\OperatorTok{-}\StringTok{ }\KeywordTok{cos}\NormalTok{(e }\OperatorTok{-}\StringTok{ }\NormalTok{beta)   }\CommentTok{## - cos(e-b), right}
\NormalTok{    tdl <-}\StringTok{ }\KeywordTok{pmax}\NormalTok{(p }\OperatorTok{-}\StringTok{ }\FloatTok{1.28} \OperatorTok{*}\StringTok{ }\KeywordTok{abs}\NormalTok{(Cl), }\DecValTok{0}\NormalTok{)  }\CommentTok{## time delays, left}
\NormalTok{    tdr <-}\StringTok{ }\KeywordTok{pmax}\NormalTok{(p }\OperatorTok{-}\StringTok{ }\FloatTok{1.28} \OperatorTok{*}\StringTok{ }\KeywordTok{abs}\NormalTok{(Cr), }\DecValTok{0}\NormalTok{)  }\CommentTok{## time delays, right}
\NormalTok{    A <-}\StringTok{ }\KeywordTok{sign}\NormalTok{(Cl)               }\CommentTok{## measurement outcomes, left}
\NormalTok{    B <-}\StringTok{ }\KeywordTok{sign}\NormalTok{(Cr)               }\CommentTok{## measurement outcomes, right}
\NormalTok{    AB <-}\StringTok{ }\NormalTok{A }\OperatorTok{*}\StringTok{ }\NormalTok{B                 }\CommentTok{## product of outcomes}
\NormalTok{    good <-}\StringTok{ }\KeywordTok{abs}\NormalTok{(tdl}\OperatorTok{-}\NormalTok{tdr) }\OperatorTok{<}\StringTok{ }\NormalTok{coincWindow}
\NormalTok{    corrs[i] <-}\StringTok{ }\KeywordTok{mean}\NormalTok{(AB[good])}
\NormalTok{    Ns[i] <-}\StringTok{ }\KeywordTok{sum}\NormalTok{(good)}
\NormalTok{\}}
\NormalTok{corrs[K] <-}\StringTok{ }\NormalTok{corrs[}\DecValTok{1}\NormalTok{]}
\NormalTok{Ns[K] <-}\StringTok{ }\NormalTok{Ns[}\DecValTok{1}\NormalTok{]}
\KeywordTok{plot}\NormalTok{(angles }\OperatorTok{*}\StringTok{ }\DecValTok{180}\OperatorTok{/}\NormalTok{pi, corrs, }\DataTypeTok{type =} \StringTok{"l"}\NormalTok{, }\DataTypeTok{col =} \StringTok{"blue"}\NormalTok{, }
     \DataTypeTok{main =} \StringTok{"epr-clocked"}\NormalTok{, }
     \DataTypeTok{xlab =} \StringTok{"Angle (degrees)"}\NormalTok{, }\DataTypeTok{ylab =} \StringTok{"Correlation"}\NormalTok{)}
\KeywordTok{lines}\NormalTok{(angles }\OperatorTok{*}\StringTok{ }\DecValTok{180}\OperatorTok{/}\NormalTok{pi, }\OperatorTok{-}\StringTok{ }\KeywordTok{cos}\NormalTok{(angles), }\DataTypeTok{col =} \StringTok{"black"}\NormalTok{)}
\KeywordTok{legend}\NormalTok{(}\DataTypeTok{x =} \DecValTok{0}\NormalTok{, }\DataTypeTok{y =} \FloatTok{1.0}\NormalTok{, }\DataTypeTok{legend =} \KeywordTok{c}\NormalTok{(}\StringTok{"epr-clocked"}\NormalTok{, }\StringTok{"neg cosine"}\NormalTok{), }
       \DataTypeTok{text.col =} \KeywordTok{c}\NormalTok{(}\StringTok{"blue"}\NormalTok{, }\StringTok{"black"}\NormalTok{), }\DataTypeTok{lty =} \DecValTok{1}\NormalTok{, }\DataTypeTok{col =} \KeywordTok{c}\NormalTok{(}\StringTok{"blue"}\NormalTok{, }\StringTok{"black"}\NormalTok{))}
\KeywordTok{plot}\NormalTok{(angles }\OperatorTok{*}\StringTok{ }\DecValTok{180}\OperatorTok{/}\NormalTok{pi, Ns }\OperatorTok{/}\StringTok{ }\NormalTok{M, }\DataTypeTok{type =} \StringTok{"l"}\NormalTok{, }\DataTypeTok{col =} \StringTok{"blue"}\NormalTok{, }
     \DataTypeTok{xlab =} \StringTok{"Angle (degrees)"}\NormalTok{,}
     \DataTypeTok{main =} \StringTok{"Rate of accepted particle pairs"}\NormalTok{, }\DataTypeTok{ylim =} \KeywordTok{c}\NormalTok{(}\DecValTok{0}\NormalTok{, }\DecValTok{1}\NormalTok{))}
\end{Highlighting}
\end{Shaded}
\begin{figure}[H]
    \centering
    \begin{minipage}{0.5\textwidth}
        \centering
        \includegraphics[width=0.95\textwidth]{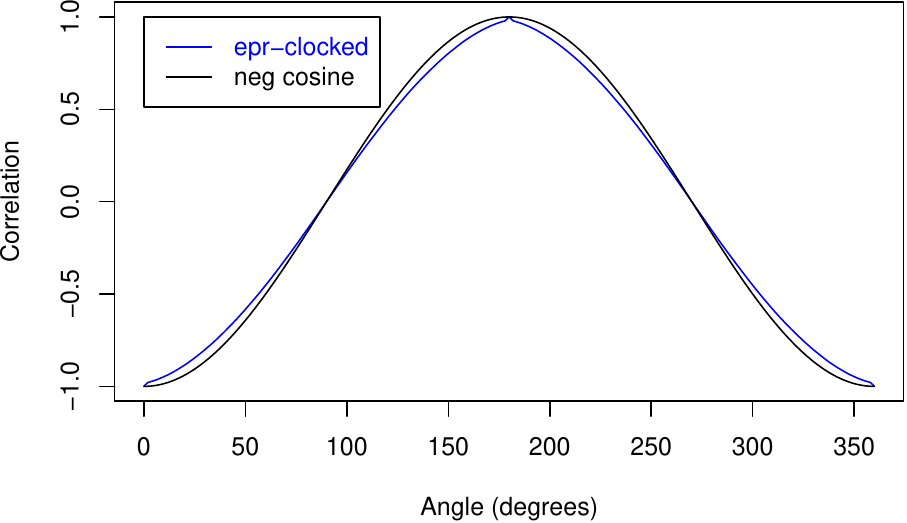} 
        \caption{epr-clocked}
    \end{minipage}\hfill
    \begin{minipage}{0.5\textwidth}
        \centering
        \includegraphics[width=0.95\textwidth]{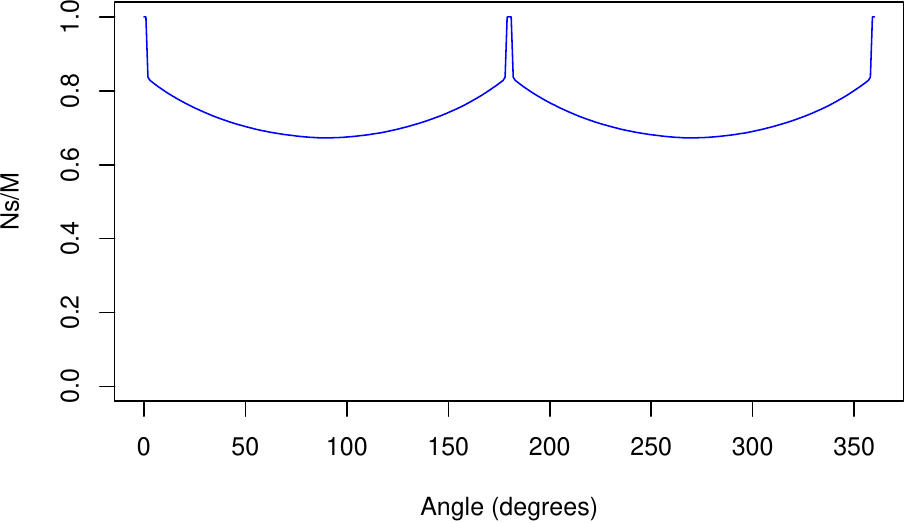} 
        \caption{Rate of detected particle pairs}
    \end{minipage}
\end{figure}

The results are almost identical. The model is rather simple. 
There are just two hidden variables: a uniformly distributed angle \(E\) in
\([0, 2\pi]\) and independently thereof, a random number
\(P = 4 \sin^2(\Theta)\) in \([0, 1]\) where \(\Theta\) is uniformly
distributed in \([0, \pi/6]\).

At Alice's measurement device, where the setting is \(\alpha\), the
measurement outcome is \(\text{sign}\cos(E - \alpha)\). At Bob's
measurement device, where the setting is \(\beta\), the measurement
outcome is \(\text{sign}\cos(E - \beta + \pi)\).

During measurement, the particles experience time delays. Alice's
particle's time delay is \(\max(P - 1.28 |\cos(E - \alpha)|, 0)\) and
Bob's is \(\max(P - 1.28 |\cos(E - \beta)|, 0)\). Notice that if
\(\alpha = \beta\) or if \(\alpha = \beta + \pi\), the time delays of
the two particles are identically equal to one another.

Finally, the two detections are accepted as belonging to one particle
pair if the difference between their two delay times is less than 0.034;
i.e., if they are detected within the same time interval of length
maximally 0.034.

The full \lq\lq\texttt{epr-clocked}\rq\rq\, model adds on top of this
simple core, some further (relatively small) sources of noise, which do
serve to smooth out the anomalous spike when the two particles are
measured in the same direction. Moreover, in his simulations, Michel
Fodje does not measure at fixed directions, but samples measurement
directions uniformly at random in the circle. When he computes
correlations, he has to bin measurement directions, resulting in another
source of noise, again further smoothing out anomalous features in the
observed correlation curve.

\hypertarget{appendix-epr-clocked-optimized}{%
\section{Appendix: epr-clocked
optimized}\label{appendix-epr-clocked-optimized}}

Inspection of the data-sets generated by the last model turned up
something very interesting: the coincidence window is so small, that for
most measurement settings, a pair of detections is only accepted as a
pair if both particles experience the same, zero, time delay. But this
means that the two detections are accepted as a pair if and only if
\(P < 1.28 |\cos(E - \alpha)\) and \(P < 1.28 |\cos(E - \beta)|\),
because only in this case are both time delays identically equal to
zero.

Now we are free to divide throughout in these inequalities by the
constant 1.28 and absorb it into the random variable \(P\), and we are
free to pick a different distribution for \(P\). So let us take \(P\)
the same as in the Pearle model. Let us also reduce the size of the
coincidence window to make it almost impossible for particles which
experience non-zero time delays to become paired with their partners.

\begin{Shaded}
\begin{Highlighting}[]
\NormalTok{coincWindow <-}\StringTok{ }\FloatTok{0.000001}
\NormalTok{p <-}\StringTok{ }\DecValTok{2}\OperatorTok{/}\KeywordTok{sqrt}\NormalTok{(}\DecValTok{1} \OperatorTok{+}\StringTok{ }\DecValTok{3} \OperatorTok{*}\StringTok{ }\NormalTok{U) }\OperatorTok{-}\StringTok{ }\DecValTok{1}  \CommentTok{## Pearle's choice}
\CommentTok{## Loop through measurement vectors 'a' }
\CommentTok{##          (except last = 360 degrees = first)}
\ControlFlowTok{for}\NormalTok{ (i }\ControlFlowTok{in} \DecValTok{1}\OperatorTok{:}\NormalTok{(K }\OperatorTok{-}\StringTok{ }\DecValTok{1}\NormalTok{)) \{}
\NormalTok{    alpha <-}\StringTok{ }\NormalTok{angles[i]}
\NormalTok{    Cl <-}\StringTok{ }\KeywordTok{cos}\NormalTok{(e }\OperatorTok{-}\StringTok{ }\NormalTok{alpha)    }\CommentTok{## cos(e-a), left}
\NormalTok{    Cr <-}\StringTok{ }\OperatorTok{-}\StringTok{ }\KeywordTok{cos}\NormalTok{(e }\OperatorTok{-}\StringTok{ }\NormalTok{beta)   }\CommentTok{## - cos(e-b), right}
\NormalTok{    tdl <-}\StringTok{ }\KeywordTok{pmax}\NormalTok{(p }\OperatorTok{-}\StringTok{ }\KeywordTok{abs}\NormalTok{(Cl), }\DecValTok{0}\NormalTok{)  }\CommentTok{## time delays, left}
\NormalTok{    tdr <-}\StringTok{ }\KeywordTok{pmax}\NormalTok{(p }\OperatorTok{-}\StringTok{ }\KeywordTok{abs}\NormalTok{(Cr), }\DecValTok{0}\NormalTok{)  }\CommentTok{## time delays, right}
\NormalTok{    A <-}\StringTok{ }\KeywordTok{sign}\NormalTok{(Cl)               }\CommentTok{## measurement outcomes, left}
\NormalTok{    B <-}\StringTok{ }\KeywordTok{sign}\NormalTok{(Cr)               }\CommentTok{## measurement outcomes, right}
\NormalTok{    AB <-}\StringTok{ }\NormalTok{A }\OperatorTok{*}\StringTok{ }\NormalTok{B                 }\CommentTok{## product of outcomes}
\NormalTok{    good <-}\StringTok{ }\KeywordTok{abs}\NormalTok{(tdl}\OperatorTok{-}\NormalTok{tdr) }\OperatorTok{<}\StringTok{ }\NormalTok{coincWindow}
\NormalTok{    corrs[i] <-}\StringTok{ }\KeywordTok{mean}\NormalTok{(AB[good])}
\NormalTok{    Ns[i] <-}\StringTok{ }\KeywordTok{sum}\NormalTok{(good)}
\NormalTok{\}}
\NormalTok{corrs[K] <-}\StringTok{ }\NormalTok{corrs[}\DecValTok{1}\NormalTok{]}
\NormalTok{Ns[K] <-}\StringTok{ }\NormalTok{Ns[}\DecValTok{1}\NormalTok{]}
\KeywordTok{plot}\NormalTok{(angles }\OperatorTok{*}\StringTok{ }\DecValTok{180}\OperatorTok{/}\NormalTok{pi, corrs, }\DataTypeTok{type =} \StringTok{"l"}\NormalTok{, }\DataTypeTok{col =} \StringTok{"blue"}\NormalTok{, }
     \DataTypeTok{main =} \StringTok{"epr-clocked optimized"}\NormalTok{, }
     \DataTypeTok{xlab =} \StringTok{"Angle (degrees)"}\NormalTok{, }\DataTypeTok{ylab =} \StringTok{"Correlation"}\NormalTok{)}
\KeywordTok{lines}\NormalTok{(angles }\OperatorTok{*}\StringTok{ }\DecValTok{180}\OperatorTok{/}\NormalTok{pi, }\OperatorTok{-}\StringTok{ }\KeywordTok{cos}\NormalTok{(angles), }\DataTypeTok{col =} \StringTok{"black"}\NormalTok{)}
\KeywordTok{legend}\NormalTok{(}\DataTypeTok{x =} \DecValTok{0}\NormalTok{, }\DataTypeTok{y =} \FloatTok{1.0}\NormalTok{, }\DataTypeTok{legend =} \KeywordTok{c}\NormalTok{(}\StringTok{"epr-clock-opt"}\NormalTok{, }\StringTok{"neg cosine"}\NormalTok{), }
       \DataTypeTok{text.col =} \KeywordTok{c}\NormalTok{(}\StringTok{"blue"}\NormalTok{, }\StringTok{"black"}\NormalTok{), }\DataTypeTok{lty =} \DecValTok{1}\NormalTok{, }\DataTypeTok{col =} \KeywordTok{c}\NormalTok{(}\StringTok{"blue"}\NormalTok{, }\StringTok{"black"}\NormalTok{))}
\KeywordTok{plot}\NormalTok{(angles }\OperatorTok{*}\StringTok{ }\DecValTok{180}\OperatorTok{/}\NormalTok{pi, Ns }\OperatorTok{/}\StringTok{ }\NormalTok{M, }\DataTypeTok{type =} \StringTok{"l"}\NormalTok{, }\DataTypeTok{col =} \StringTok{"blue"}\NormalTok{, }
     \DataTypeTok{xlab =} \StringTok{"Angle (degrees)"}\NormalTok{,}
     \DataTypeTok{main =} \StringTok{"Rate of accepted particle pairs"}\NormalTok{, }\DataTypeTok{ylim =} \KeywordTok{c}\NormalTok{(}\DecValTok{0}\NormalTok{, }\DecValTok{1}\NormalTok{))}
\end{Highlighting}
\end{Shaded}
\begin{figure}[H]
    \centering
    \begin{minipage}{0.5\textwidth}
        \centering
        \includegraphics[width=0.95\textwidth]{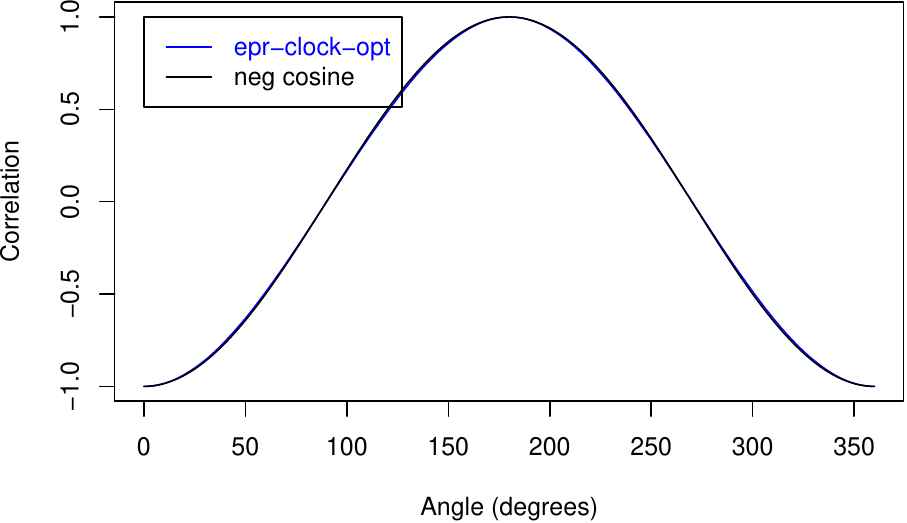} 
        \caption{epr-clocked optimized}
    \end{minipage}\hfill
    \begin{minipage}{0.5\textwidth}
        \centering
        \includegraphics[width=0.95\textwidth]{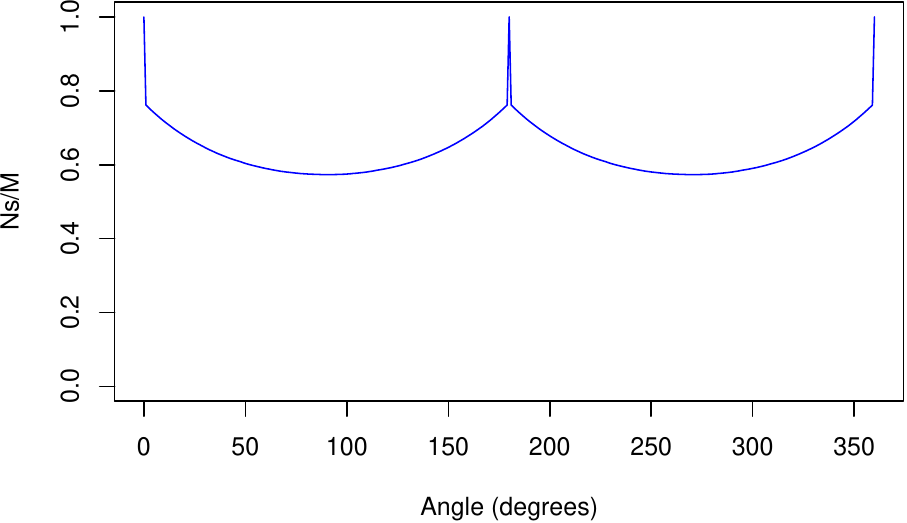} 
        \caption{Rate of detected particle pairs}
    \end{minipage}
\end{figure}

Conclusion: \texttt{epr-clocked} is actually a disguised and perturbed
version of the detection loophole model \texttt{epr-simple}. It can
therefore be slightly improved, just as \texttt{epr-simple} can be
slightly improved. However, this will not help it achieve the maximum
efficiency possible for coincidence-loophole models.

\end{document}